\documentclass{llncs}
\usepackage{graphicx}
\usepackage{subfig}
\usepackage{wrapfig}
\usepackage[ruled,vlined]{algorithm2e}
\begin{document}
	\title{Ensemble-Compression: A New Method for Parallel Training of Deep Neural Networks}
	\author{Shizhao Sun\inst{1}\thanks{This work was done when the author was visiting Microsoft Research Asia.} 
		\and Wei Chen\inst{2} 
		\and Jiang Bian\inst{2} 
		\and Xiaoguang Liu\inst{1} 
		\and Tie-Yan Liu\inst{2}}
	\institute{College of Computer and Control Engineering, Nankai University, Tianjin, P.R.China
		\and Microsoft Research, Beijing, P.R.China \\
	\email{
		sunshizhao@mail.nankai.edu.cn, 
		wche@microsoft.com,
		jiabia@microsoft.com,
		liuxg@nbjl.nankai.edu.cn,
		tyliu@microsoft.com.
		}}
	\maketitle
	\begin{abstract}
		Parallelization framework has become a necessity to speed up the training of deep neural networks (DNN) recently. Such framework typically employs the \emph{Model Average} approach, denoted as MA-DNN, in which parallel workers conduct respective training based on their own local data while the parameters of local models are periodically communicated and averaged to obtain a global model which serves as the new start of local models. However, since DNN is a highly non-convex model, averaging parameters cannot ensure that such global model can perform better than those local models. To tackle this problem, we introduce a new parallel training framework called \emph{Ensemble-Compression}, denoted as EC-DNN. In this framework, we propose to aggregate the local models by ensemble, i.e., averaging the outputs of local models instead of the parameters. As most of prevalent loss functions are convex to the output of DNN, the performance of ensemble-based global model is guaranteed to be at least as good as the average performance of local models. However, a big challenge lies in the explosion of model size since each round of ensemble can give rise to multiple times size increment. Thus, we carry out model compression after each ensemble, specialized by a distillation based method in this paper, to reduce the size of the global model to be the same as the local ones. Our experimental results demonstrate the prominent advantage of EC-DNN over MA-DNN in terms of both accuracy and speedup.
	\end{abstract}
	\begin{keywords}
		Parallel machine learning, Distributed machine learning, Deep learning, Ensemble method.
	\end{keywords}
	\section{Introduction}\label{sec:intro}
	Recent rapid development of deep neural networks (DNN) has demonstrated that its great success mainly comes from big data and big models \cite{szegedy2014going,he2015delving}. However, it is extremely time-consuming to train a large-scale DNN model over big data. To accelerate the training of DNN, parallelization frameworks like MapReduce \cite{dean2008mapreduce} and Parameter Server\cite{li2014scaling,dean2012large} have been widely used. A typical parallel training procedure for DNN consists of continuous iterations of the following three steps. First, each worker trains the local model based on its own local data by stochastic gradient decent (SGD) or any of its variants. Second, the parameters of the local DNN models are communicated and aggregated to obtain a global model, e.g., by averaging the identical parameter of each local models \cite{zhang2014improving,povey2014parallel}. Finally, the obtained global model is used as a new starting point of the next round of local training. We refer the method that performs the aggregation by averaging model parameters as \emph{MA}, and the corresponding parallel implementation of DNN as \emph{MA-DNN}.
	
	However, since DNN is a highly non-convex model, the loss of the global model produced by MA cannot be guaranteed to be upper bounded by the average loss of the local models. In other words, the global model obtained by MA-DNN might even perform worse than any local model, especially when the local models fall into different local-convex domains. As the global model will be used as a new starting point of the successive iterations of local training, the poor performance of the global model will drastically slow down the convergence of the training process and further hurt the performance of the final model.
	
	To tackle this problem, we propose a novel framework for parallel DNN training, called \emph{Ensemble-Compression (EC-DNN)}, the key idea of which is to produce the global model by ensemble instead of MA. Specifically, the ensemble method aggregates local models by averaging their outputs rather than their parameters. Equivalently, the global model produced by ensemble is a larger network with one additional layer which takes the outputs of local models as inputs with weights as $1/K$, where $K$ is the number of local models. Since most of widely-used loss functions for DNN (e.g., cross entropy loss, square loss, and hinge loss) are convex with respect to the output vector of the model, the loss of the global model produced by ensemble can be upper bounded by the average loss of the local models. Empirical evidence in~\cite{szegedy2014going,ciresan2012multi} even show that the ensemble model of DNN, i.e., the global model, is usually better than any base model, i.e., the local model. According to previous theoretical and empirical studies~\cite{kuncheva2003ensemble,sollich1996ensemble}, ensemble model tend to yield better results when there exists a significant diversity among local models. Therefore, we train the local models for a longer period for EC-DNN to increase the diversity among them. In other words, EC-DNN yields less communication frequency than MA-DNN, which further emphasizes the advantages of EC-DNN by reducing communication cost as well as increasing robustness to limited network bandwidth.
	
	There is, however, no free lunch. In particular, the ensemble method will critically increase the model complexity: the resultant global model with one additional layer will be $K$ times wider than any of the local models. Several ensemble iterations may result in explosion of the size of the global model. To address this problem, we further propose an additional compression step after the ensemble. This approach cannot only restrict the size of the resultant global model to be the same size as local ones, but also preserves the advantage of ensemble over MA. Given that both ensemble and compression steps are dispensable in our new parallelization framework, we name this framework as EC-DNN. As a specialization of the EC-DNN framework, we adopt the distillation based compression \cite{bucilu2006model,romero2014fitnets,hinton2015distilling}, which produces model compression by distilling the predictions of big models. Nevertheless, such distillation method requires extra time for training the compressed model. To tackle this problem, we seek to integrate the model compression into the process of local training by designing a new combination loss, a weighted interpolation between the loss based on the pseudo labels produced by global model and that based on the true labels. By optimizing such combination loss, we can achieve model compression in the meantime of local training.
	
	We conducted comprehensive experiments on CIFAR-10, CIFAR-100, and ImageNet datasets, w.r.t. different numbers of local workers and communication frequencies. The experimental results illustrate a couple of important observations: 1) Ensemble is a better model aggregation method than MA consistently. MA suffers from that the performance of the global model could vary drastically and even be much worse than the local models; meanwhile, the global model obtained by the ensemble method can consistently outperform the local models. 2) In terms of the end-to-end results, EC-DNN stably achieved better test accuracy than MA-DNN in all the settings. 3) EC-DNN can achieve better performance than MA-DNN even when EC-DNN communicates less frequently than MA-DNN, which emphasizes the advantage of EC-DNN in training a large-scale DNN model as it can significantly reduce the communication cost.
	\section{Preliminary: Parallel Training of DNN}\label{sec:preliminary}
	In the following of this paper, we denote a DNN model as $f(\mathbf{w})$ where $\mathbf{w}$ represents the parameters of this DNN model.	In addition, we denote the outputs of the model $f(\mathbf{w})$ on the input $x$ as $f(\mathbf{w};x)=(f(\mathbf{w};x,1),\dots,f(\mathbf{w};x,C))$, where $C$ is the number of classes and $f(\mathbf{w};x,c)$ denotes the output (i.e., the score) for the $c$-th class. DNN is a highly non-convex model due to the non-linear activations and poolings after many layer.
	
	In the parallel training of DNN, suppose that there are $K$ workers and each of them holds a local dataset $D_k=\{(x_{k,1},y_{k,1}),\dots, (x_{k,m_k},y_{k,m_k})\}$ with size $m_k$, $k\in\{1,\dots,K\}$. Denote the weights of the DNN model at the iteration $t$ on the worker $k$ as $\mathbf{w}_k^t$. The communication between the workers is invoked after every $\tau$ iterations of updates for the weights, and we call $\tau$ the communication frequency. A typical parallel training procedure for DNN implements the following three steps in an iterative manner until the training curve converges. 
	
	\noindent\emph{1. Local training: }At iteration $t$, worker $k$ updates its local model by using SGD. Such local model is updated for $\tau$ iterations before the cross-machine synchronization. 
	
	\noindent\emph{2. Model aggregation: }The parameters of local models are communicated across machines. Then, a global model is produced by aggregating local models according to certain aggregation method. 
	
	\noindent\emph{3. Local model reset: }The global model is sent back to the local workers, and set as the starting point for the next round of local training.
	
	We denote the aggregation method in the second step as $G(\mathbf{w}_1^t,\dots,\mathbf{w}_K^t)$ and the weights of the global model as $\bar{\mathbf{w}}^{t}$. That is, $f(\bar{\mathbf{w}}^{t})=G(\mathbf{w}_1^t,\dots,\mathbf{w}_K^t)$, where $t=\tau,2\tau,\cdots$. A widely-used aggregation method is \emph{model average (MA)}, which averages each parameter over all the local models, i.e.,
	\begin{small}
		\begin{equation}
		G_{MA}\left(\mathbf{w}_1^t,\dots,\mathbf{w}_K^t\right)= f\left(\frac{1}{K}\sum_{k=1}^{K}\mathbf{w}_{k}^{t}\right), t=\tau, 2\tau,\cdots.
		\end{equation}
	\end{small}We denote the parallel training method of DNN that using MA as MA-DNN for ease of reference. 
	
	With the growing efforts in parallel training for DNN, many previous studies~\cite{zhang2014improving,povey2014parallel,dean2012large,zhang2015deep,chen2016revisiting,chen2016scalable} have paid attention to MA-DNN. NG-SGD~\cite{povey2014parallel} proposes an approximate and efficient implementation of Natural Gradient for SGD (NG-SGD) to improve the performance of MA-DNN. EASGD~\cite{zhang2015deep} improves MA-DNN by adding an elastic force which links the weights of the local models with the weights of the global model. BMUF~\cite{chen2016scalable} leverages data parallelism and blockwise model-update filtering to improve the speedup of MA-DNN. All these methods aim at solving different problems with us, and our method can be used with those methods simultaneously.  
	
	\section{Model Aggregation: MA vs. Ensemble}\label{sec:ma_vs_ec}
	In this section, we first reveal why the MA method cannot guarantee to produce a global model with better performance than local models. Then, we propose to use ensemble method to perform the model aggregation, which in contrast can ensure to perform better over local models.
	
	MA was originally proposed for convex optimization. If the model is convex w.r.t. the parameters and the loss is convex w.r.t. the model outputs, the performance of the global model produced by MA can be guaranteed to be no worse than the average performance of local models. This is because, when $f(\cdot)$ is a convex model, we have,
	\begin{small}
		\begin{equation}
		\label{eqn:pdnn_model_convex}
		\mathcal{L}\left(f\left(\bar{\mathbf{w}}^{t};x\right),y\right)=
		\mathcal{L}\left(f\left(\frac{1}{K}\sum_{k=1}^K \mathbf{w}_k^{t};x\right),y\right) \le \mathcal{L}\left(\frac{1}{K}\sum_{k=1}^{K} f\left(\mathbf{w}_k^{t};x\right),y\right).
		\end{equation}
	\end{small}Moreover, when the loss is also convex w.r.t. the model outputs $f(\cdot;x)$, we have,
	\begin{small}
		\begin{equation}
		\label{eqn:pdnn_output_convex}
		\mathcal{L}\left(\frac{1}{K}\sum_{k=1}^{K}f\left(\mathbf{w}_k^t;x\right),y\right)\le \frac{1}{K}\sum_{k=1}^{K}\mathcal{L}\left(f\left(\mathbf{w}_k^t;x\right),y\right).
		\end{equation}
	\end{small}By combining inequalities (\ref{eqn:pdnn_model_convex}) and (\ref{eqn:pdnn_output_convex}), we can see that it is quite effective to apply MA in the context of convex optimization, since the loss of the global model by MA is no greater than the average loss of local models in such context. 
	
	However, DNN is indeed a highly non-convex model due to the existence of activation functions and pooling functions (for convolutional layers). Therefore, the above properties of MA for convex optimization does not hold for DNN such that the MA method cannot produce any global model with guaranteed better performance than local ones. Especially, when the local models are in the neighborhoods of different local optima, the global model based on MA could be even worse than any of the local models. Furthermore, given that the global model is usually used as the starting point of the next round of local training, the performance of the final model could hardly be good if a global model in any round fails to achieve good performance. Beyond the theoretical analysis above, the experimental results reported in Section~\ref{subsec:psgd_ec} and previous studies~\cite{povey2014parallel,chen2016scalable} also revealed such problem.
	
	While the DNN model itself is non-convex, we notice that most of widely-used loss functions for DNN are convex w.r.t. the model outputs (e.g., cross entropy loss, square loss, and hinge loss). Therefore, Eq.(\ref{eqn:pdnn_output_convex}) holds,  which indicates that averaging the output of the local models instead of their parameters guarantees to yield better performance than local models. To this end, we propose to \emph{ensemble} the local models by averaging their outputs as follows,
	\begin{small}
		\begin{equation}\label{eqn:ensemble}
		G_E\left(\mathbf{w}_1^t,\dots,\mathbf{w}_K^t\right)=\frac{1}{K}\sum_{k=1}^{K}f\left(\mathbf{w}_{k}^t;x\right), t=\tau,2\tau,\cdots.
		\end{equation}
	\end{small}
	\section{EC-DNN}\label{sec: ec_alg}
	In this section, we first introduce the EC-DNN framework, which employs ensemble for model aggregation. Then, we introduce a specific implementation of EC-DNN that adopts distillation for the compression. At last, we take some further discussions for the time complexity of EC-DNN and the comparison with traditional ensemble methods.
	\subsection{Framework}		
	The details of EC-DNN framework is shown in Alg.~\ref{alg:ensemble_compression}. Note that, in this paper, we focus on the synchronous case\footnote{As shown in \cite{chen2016revisiting}, MA-DNN in synchronous case converges faster and achieves better test accuracy than that in asynchronous case.} within the MapReduce framework, but EC-DNN can be generalized into the asynchronous case and parameter server framework as well. Similar to other popular parallel training methods for DNN, EC-DNN iteratively conducts local training, model aggregation, and local model reset. 
	
	\emph{1. Local training: }The local training process of EC-DNN is the same as that of MA-DNN, in which the local model is updated by SGD. Specifically, at iteration $t$, worker $k$ updates its local model from $\mathbf{w}_k^t$ to $\mathbf{w}_k^{t+1}$ by minimizing the training loss using SGD, i.e., $\mathbf{w}_k^{t+1}=\mathbf{w}_k^t-\eta\Delta(\mathcal{L}(f(\mathbf{w}_k^t;x_k),y_k))$, where $\eta$ is the learning rate, and $\Delta(\mathcal{L}(f(\mathbf{w}_k^t;x_k),y_k))$ is the gradients of the empirical loss $\mathcal{L}(f(\mathbf{w}_k^t;x_k),y_k)$ of the local model $f(\mathbf{w}_k^t)$ on one mini batch of the local dataset $D_k$. One local model will be updated for $\tau$ iterations before the cross-machine synchronization. 
	
	\emph{2. Model aggregation: }The goal of model aggregation is to communicate the parameters of local models, i.e., $\mathbf{w}_1^{t}\dots \mathbf{w}_K^t$, across machines. To this end, a global model is produced by ensemble according to $G_E(\mathbf{w}_1^t,\dots,\mathbf{w}_K^t)$ in Eq.(\ref{eqn:ensemble}), i.e., averaging the outputs of the local models. Equivalently, the global model produced by ensemble is a larger network with one additional layer, whose outputs consist of $C$ nodes representing $C$ classes, and whose inputs are those outputs from local models with weights as $1/K$, where $K$ is the number of local models. Therefore, the weights of global model $\bar{\mathbf{w}}^t$ can be denoted as $\bar{\mathbf{w}}^t=\{\mathbf{w}^t_1,\dots,\mathbf{w}^t_K,\frac{1}{K}\}$, $t=\tau,2\tau,\dots$ 
	
	Note that such ensemble-produced global model (i.e., $f(\bar{\mathbf{w}_t})$) is one layer deeper and $K$ times wider than the local model (i.e., $f(\mathbf{w}_t^k)$). Therefore, continuous rounds of ensemble process will easily give rise to a global model with exploding size. To avoid this problem, we propose introducing a compression process (i.e., Compression($\mathbf{w}_k^t,\bar{\mathbf{w}}^t,D_k$) in Alg.~\ref{alg:ensemble_compression}) after ensemble process, to compress the resultant global model to be the same size as those local models while preserving the advantage of the ensemble over MA. We denote the compressed model for the global model $\bar{\mathbf{w}}_t$ on the worker $k$ as $\tilde{\mathbf{w}}_k^t$. 
	
		\begin{wrapfigure}{r}{0.55\textwidth}
			\begin{small}
				\begin{algorithm}[H]
					\caption{EC-DNN($D_k$)}
					\label{alg:ensemble_compression}
					Randomly initialize $\mathbf{w}_k^0$ and set $t=0$\;
					\While{stop criteria are not satisfied}{
						$\mathbf{w}_k^{t+1}\gets \mathbf{w}_k^t-\eta\Delta(\mathcal{L}(f(\mathbf{w}_k^t;x_k),y_k))$\;
						$t\gets t+1$\; 
						\If{$\tau$ divides $t$}{ 
							$\bar{\mathbf{w}}^t\gets\{\mathbf{w}_1^t,\dots,\mathbf{w}_K^t,\frac{1}{K}\}$\;
							$\tilde{\mathbf{w}}_k^{t}\gets$ Compression($\mathbf{w}_k^t, \bar{\mathbf{w}}^t, D_k$)\;
							$\mathbf{\mathbf{w}}_k^{t}\gets \tilde{\mathbf{\mathbf{w}}}_k^{t}$.
						}
					}
					\Return ${\mathbf{w}}_k^t$
				\end{algorithm}
			\end{small}
		\end{wrapfigure}	
	\emph{3. Local model reset: }The compressed model will be set as the new starting point of the next round of local training, i.e., $\mathbf{w}_k^{t}=\tilde{\mathbf{w}}^{t}$ where $t=\tau,2\tau,\cdots$.
	
	At the end of the training process, EC-DNN will output $K$ local models and we choose the model with the smallest training loss as the final one. Note that, we can also take the global model (i.e., the ensemble of $K$ local models) as the final model if there are enough computation and storage resources for the test.
	
	\subsection{Implementations}\label{subsec:implementation}
	Algorithm~\ref{alg:ensemble_compression} contains two sub-problems that need to be addressed more concretely: 1) how to train local models that can benefit more to the ensemble model; 2) how to compress the global model without costing too much extra time. 
	
	\subsubsection{Diversity Driven Local Training.} In order to improve the performance of ensemble, it is necessary to generate diverse local models other than merely accurate ones~\cite{kuncheva2003ensemble,sollich1996ensemble}. Therefore, in the local training phase, i.e., the third line in Alg.~\ref{alg:ensemble_compression}, we minimize both the loss on training data and the similarity between the local models, which we call \emph{diversity regularized local training loss}. For the $k$-th worker, it is defined as follows,
	\begin{small}
		\begin{equation}\label{eqn:ensemble_loss}
		\mathcal{L}_\mathrm{LS }^k(f(\mathbf{w}_k;x_{k,i}),y_{k,i})=\sum_{i=1}^{m_k}\left(\mathcal{L}\left(f(\mathbf{w}_k;x_{k,i}),y_{k,i}\right)+\alpha \mathcal{L}_\mathrm{sim}\left(f\left(\mathbf{w}_k;{x_{k,i}}\right),\bar{z}_{k,i}\right)\right),
		\end{equation}
	\end{small}where $\bar{z}_{y,i}$ is the average of the outputs of the latest compressed models for input $x_{k,i}$. In our experiments, the local training loss $\mathcal{L}$ takes the form of cross entropy, and the similarity loss $\mathcal{L}_\mathrm{sim}$ takes the form of $-l_2$ distance. The smaller $\mathcal{L}_\mathrm{sim}$ is, the farther the outputs of a local model is from the average of outputs of the latest compressed models, and hence  the farther (or the more diverse) the local models are from (or with) each other.
	
	\subsubsection{Distillation Based Compression.}
	In order to compress the global model to the one with the same size as the local model, we use distillation base compression method\footnote{Other algorithms for the compression \cite{chen2015compressing,gong2014compressing,denil2013predicting,denton2014exploiting,rigamonti2013learning,han2015learning} can also be used for the same purpose, but different techniques may be required in order to plug these compression algorithms into the EC-DNN framework.}~\cite{bucilu2006model,romero2014fitnets,hinton2015distilling}, which obtains a compressed model by letting it mimic the predictions of the global model. In order to save the time for compression, in compression process, we minimize the weighted combination of the local training loss and the pure compression loss, which we call \emph{accelerated compression loss}. For the $k$-th worker, it is defined as follows,
	\begin{small}
		\begin{equation}\label{eqn:compress_loss}
		\mathcal{L}_\mathrm{LC}^k(f(\mathbf{w}_k;x_{k,i}),y_{k,i})= \sum_{i=1}^{m_k}\left(\mathcal{L}\left(f\left(\mathbf{w}_k; x_{k,i}\right), y_{k,i}\right) + \beta \mathcal{L}_{\mathrm{comp}}\left(f\left(\mathbf{w}_k; x_{k,i}\right),\bar{y}_{k,i}\right)\right), 
		\end{equation}
	\end{small}where $\bar{y}_{k,i}$ is the output of the latest ensemble model for the input $x_{k,i}$. In our experiments, the local training loss $\mathcal{L}$ and the pure compression loss $\mathcal{L}_{\mathrm{comp}}$ both take the form of cross entropy loss. By reducing the loss between $f\left(\mathbf{w}_k; x_{k,i}\right)$ and the pseudo labels $\{\bar{y}_{k,i}; i\in[m_k] \}$, the compressed model can play the similar function as the ensemble model. 
	
	We denote the distillation based compression process as Compression$_\mathrm{{distill}}$ ($\mathbf{w}_k^t$, $\bar{\mathbf{w}}^t$, $D_k$), and show its details in  Alg.~\ref{alg:compression}. First, on the $k$-th local worker, we construct a new training data $\hat{D}_k$ by relabeling the original dataset $D_k$ with the pseudo labels produced by the global model, i.e., $\{\bar{y}_{k,i},i\in[m_k]\}$. Specifically, when producing pseudo labels, we first produce the predictions of each local models respectively, and then average the predictions of all the local models. By this way, we keep using the same amount of GPU memory as MA-DNN throughout the training, because the big global model, which is $K$ times larger than the local model, has never been established in GPU memory. Then, we optimize the accelerated compression loss $\mathcal{L}_\mathrm{LC}^k$ in Eq.(\ref{eqn:compress_loss}) by SGD for $p$ iterations. We initialize the parameters of the compressed model $\tilde{\mathbf{w}}_k^t$ by the parameters of the latest local model $\mathbf{w}_k^t$ instead of random numbers. Finally, the obtained compressed model $\tilde{\mathbf{w}}_k^{t+p}$ is returned, which will be set as the new starting point of next round of the local training.
	
	\begin{wrapfigure}{r}{0.64\textwidth}
		\begin{small}
			\begin{algorithm}[H]
				\caption{Compression$_\mathrm{{distill}}$($\mathbf{w}_k^t, \bar{\mathbf{w}}_k^t, D_k$)}
				\label{alg:compression}
				\For{$j\in[m_k]$}{
					\For{$c \in [C]$}{
						$\bar{y}_{k,j,c} \gets \frac{1}{K}\sum_{r=1}^{K}f(\mathbf{w}_r^t;x_{k,j},c)$\;
					}
					$\bar{y}_{k,j}=(\bar{y}_{k,j,1},\dots,\bar{y}_{k,j,C})$\;
				}
				$\hat{D}_k\gets\{(x_{k,1},y_{k,1},\bar{y}_{k,1}),\dots,(x_{k,m_k},y_{k,m_k},\bar{y}_{k,m_k})\}$\;
				Set $\tilde{\mathbf{w}}_k^{t}=\mathbf{w}_k^{t}$ and $i=0$\;
				\While {$i\le p$}{
					$\tilde{\mathbf{w}}_k^{t+i+1}\gets \tilde{\mathbf{w}}_k^{t+i}-\eta\Delta(\mathcal{L}_{\mathrm{LC}}^k(f(\tilde{\mathbf{w}}_k^{t+i};x_k),y_k))$\;
					$i\gets i+1$\;
				}
				\Return $\tilde{\mathbf{w}}_k^{t+p}$.
			\end{algorithm}
		\end{small}
	\end{wrapfigure}	
	We can find that minimizing the diversity regularized loss $\mathcal{L}_\mathrm{LS}^k$ for local training (Eq.(\ref{eqn:ensemble_loss})) and minimizing the accelerated  compression loss $\mathcal{L}_\mathrm{LC}^k$ for compression (Eq.(\ref{eqn:compress_loss})) are two opposite but complementary tasks. They need to leverage information generated by each other into their own optimization. Specifically, the local training phase leverages $\bar{z}_{k,i}$ based on compressed model while the compression process uses $\bar{y}_{k,i}$ provided by local models. Due to such structural duality, we take advantage of a new optimization framework, i.e. dual learning~\cite{he2016dual}, to improve the performance of both tasks simultaneously.
	
	\subsection{Time Complexity}\label{subsec:time}
	We compare the time complexity of MA-DNN and EC-DNN from two aspects:
	
	\emph{1. Communication time: }Parallel DNN training process is usually sensitive to the communication frequency $\tau$. Different parallelization frameworks yield various optimal $\tau$. In particular, EC-DNN prefers larger $\tau$ compared to MA-DNN. Essentially, less frequent communication across workers can give rise to more diverse local models, which will lead to better ensemble performance for EC-DNN.  On the other hand, much diverse local models may indicate greater probability that local models are in the neighboring of different local optima such that the global model in MA-DNN is more likely to perform worse than local ones. The poor performance of the global model will significantly slow down the convergence and harm the performance of the final model. Therefore, EC-DNN yields less communication time than MA-DNN.
	
	\emph{2. Computational time: } According to the analysis in Sec~\ref{subsec:implementation}, EC-DNN does not consume extra computation time for model compression since the compression process has been integrated into the local training phase, as shown in Eq.(\ref{eqn:compress_loss}). Therefore, compared with MA-DNN, EC-DNN only requires additional time to relabel the local data using the global model, which approximately equals to the maximal time of the feed-forward propagation over the local dataset.  We call such extra time ``\emph{relabeling time}" for ease of reference. To limit the relabeling time on large datasets, we choose to relabel a portion of the local data, denoted as $\mu$. Our experimental results in Section~\ref{subsec:psgd_ec} will demonstrate that the relabeling time can be controlled within a very small amount compared to the training time of DNN. Therefore, EC-DNN can cost only a slightly more or roughly equal computational time over MA-DNN.
	
	Overall, compared to MA-DNN, EC-DNN is essentially more time-efficient as it can reduce the communication cost without significantly increasing computational time.  
	
	\subsection{Comparison with Traditional Ensemble Methods}\label{subsec:comparison_ensemble}
	Traditional ensemble methods for DNN~\cite{szegedy2014going,ciresan2012multi} usually first train several DNN models independently without communication and make ensemble of them in the end.  We denote such method as E-DNN. E-DNN was proposed to improve the accuracy of DNN models by reducing variance and it has no necessity to train base models with parallelization framework. In contrast, EC-DNN is a parallel algorithm aiming at training DNN models faster without the loss of the accuracy by leveraging a cluster of machines.
	
	Although E-DNN can be viewed as a special case of EC-DNN with only one final communication and no compression process, the intermediate communications in EC-DNN will make it outperform E-DNN. The reasons are as follows: 1) local workers has different local data, the communications during the training will help local models to be consensus towards the whole training data; 2) the local models of EC-DNN can be continuously optimized by compressing the ensemble model after each ensemble process. Then, another round of ensemble will result in more advantage for EC-DNN over E-DNN since the local models of EC-DNN has been much improved. 	
	
	\section{Experiments}\label{sec:exp}
	\subsection{Experimental Setup}\label{subsec:exp_setup}	
	\paragraph{Platform.}
	Our experiments are conducted on a GPU cluster interconnected with an InfiniBand network, each machine of which is equipped with two Nvdia's K20 GPU processors. One GPU processor corresponds to one local worker. 
	
	\paragraph{Data.}
	We conducted experiments on public datasets CIFAR-10, CIFAR-100 \cite{Krizhevsky09learningmultiple} and ImageNet (ILSVRC 2015 Classification Challenge) \cite{ILSVRC15}.  
	For all the datasets, each image is normalized by subtracting the per-pixel mean computed over the whole training set. The training images are horizontally flipped but not cropped, and the test data are neither flipped nor cropped. 
	
	\paragraph{Model.}
	On CIFAR-10 and CIFAR-100, we employ NiN~\cite{lin2014network}, a 9-layer convolutional network. On ImageNet, we use GoogLeNet \cite{szegedy2014going}, a 22-layer convolutional network. We used the same tricks, including random initialization, $l_2$-regularization, Dropout, and momentum, as the original paper. All the experiments are implemented using Caffe \cite{jia2014caffe}.
	
	\paragraph{Parallel Setting.}
	On experiments on CIFAR-10 and CIFAR-100, we explore the number of workers $K \in \{4, 8\}$ and the communication frequency $\tau \in \{1, 16,$ $2000, 4000\}$ for both MA-DNN and EC-DNN. On experiments on ImageNet, we explore $K\in \{4, 8\}$ and $\tau \in \{1, 1000, 10000\}$. The communication across local workers is implemented using MPI.
	
	\paragraph{Hyperparameters Setting of EC-DNN.}
	There are four hyperparameters in EC-DNN, including 1) the coefficient of the regularization in terms of similarity between local models, i.e., $\alpha$ in Eq.(\ref{eqn:ensemble_loss}), 2) the coefficient of the model compression loss, i.e., $\beta$ in Eq.(\ref{eqn:compress_loss}), 3) the length of the compression process, i.e., $p$ in Alg.~\ref{alg:compression}, and 4) the portion of the data needed to be relabeled in the compression process $\mu$ as mentioned in Sec~\ref{subsec:time}. We tune these hyperparameters by exploring a certain range of values and then choose the one resulting in best performance. In particular, we explored $\alpha$ among $\{0.2,0.4,\dots,1\}$, and finally choose $\alpha=0.6$ on all the datasets. To decide the value of $\beta$, we explored two strategies, one of which uses consistent $\beta$ at each compression process while the other employs increasing $\beta$ after a certain percentage of compression process. In the first strategy, we explored $\beta$ among $\{0.2,0.4,\dots,1\}$, and in the second one, we explored $\beta$ among $\{0.2,0.4,\dots,1\}$, the incremental step of $\beta$ among $\{0.1,0.2\}$, and the percentage of compression process from which $\beta$ begins to increase among $\{10\%,20\%,30\%\}$. On CIFAR datasets, we choose to use $\beta=0.4$ for the first 20\% of compression processes and $\beta=0.6$ for all the other compression processes. And, on ImageNet, we choose to use consistent $\beta=1$ throughout the compression. Moreover, we explored $p$'s value among $\{5\%,10\%,\dots,20\%\}$ of the number of the mini-batches that the whole training lasts, and finally pick $p=10\%$ on all the datasets. Furthermore, we explored $\mu$'s value among $\{30\%,50\%,70\%\}$. And, we finally select $\mu=70\%$ on CIFAR datasets, and $\mu=30\%$ on ImageNet. 
	
	\subsection{Compared Methods}\label{subsec:comp_methods}
	We conduct performance comparisons on four methods:
	\begin{itemize}
		\item{S-DNN} denotes the sequential training on one GPU until convergence~\cite{lin2014network,szegedy2014going}.
		\item{E-DNN} denotes the method that trains local models independently and makes ensemble of the local models merely at the end of the training~\cite{szegedy2014going,ciresan2012multi}.
		\item{MA-DNN} refers the parallel DNN training framework with the aggregation by averaging model parameters~\cite{zhang2014improving,povey2014parallel,dean2012large,zhang2015deep,chen2016revisiting,chen2016scalable}.
		\item{EC-DNN} refers the parallel DNN training framework with the aggregation by averaging model outputs. EC-DNN applies Compression$_\mathrm{distill}$ for the compression for all the experiments in this paper.
	\end{itemize}
	Furthermore, we use EC-DNN$_L$, MA-DNN$_L$ and E-DNN$_L$ to denote the corresponding methods that take the local model with the smallest training loss as the final model, and use EC-DNN$_G$, MA-DNN$_G$ and E-DNN$_G$ to represent the respective methods that take the global model (i.e., the ensemble of local models for EC-DNN and E-DNN, and the average parameters of local models for MA-DNN) as the final model.
	\subsection{Experimental Results}\label{subsec:psgd_ec}
	\subsubsection{Model Aggregation.}
\begin{figure}[t]
		\centering
		\begin{minipage}{0.32\textwidth}
			\centering
			\subfloat[CIFAR-10]
			{
				\includegraphics[width=\textwidth]{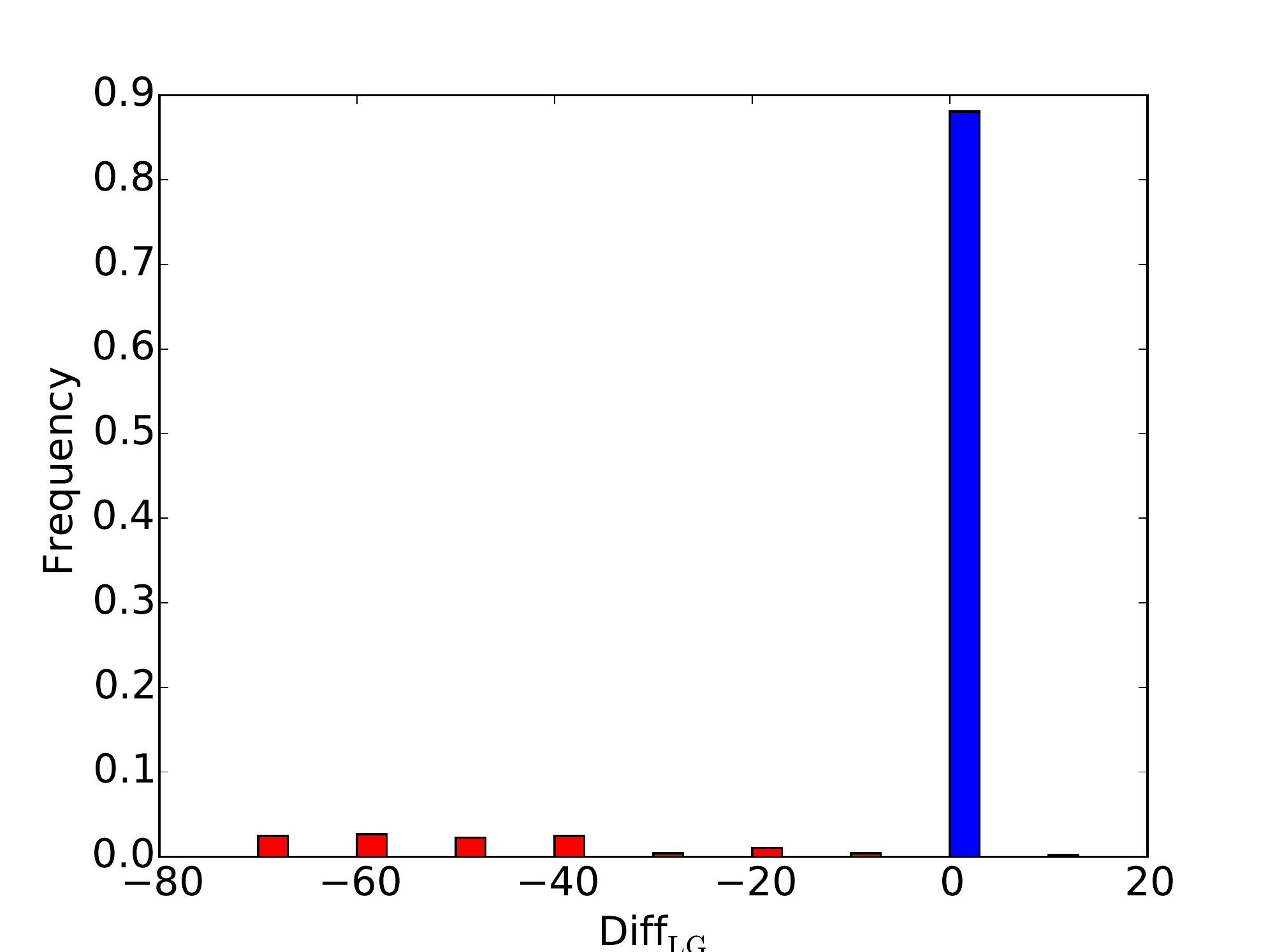}
				\label{subfig:ma_distri_cifar10}
			}
			
			\subfloat[CIFAR-100]
			{
				\includegraphics[width=\textwidth]{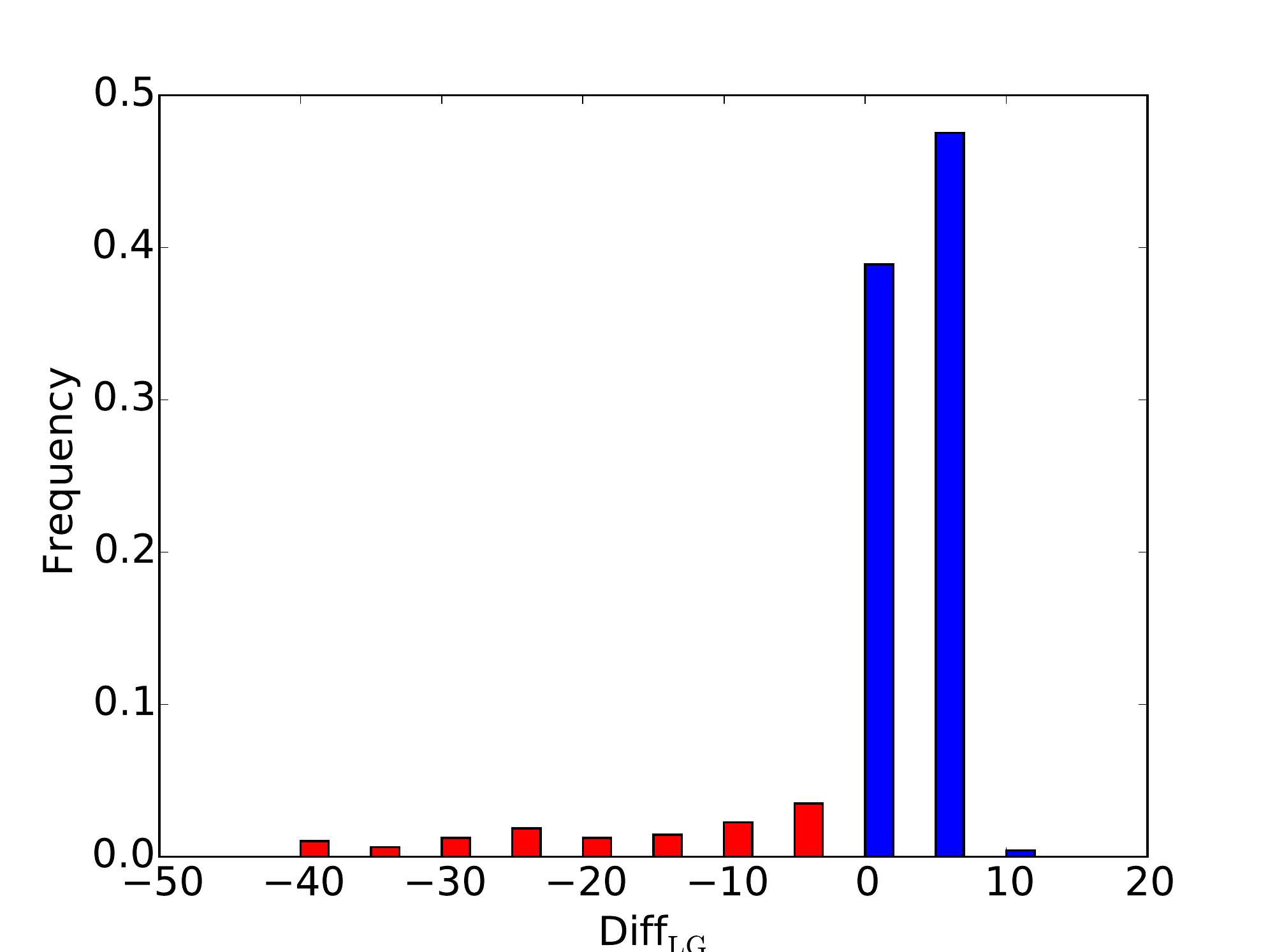}
				\label{subfig:ma_distri_cifar100}
			}
			\caption{MA.}
			\label{fig:ma_distri}
		\end{minipage}
		\begin{minipage}{0.32\textwidth}
			\centering
			\subfloat[CIFAR-10]
			{
				\includegraphics[width=\textwidth]{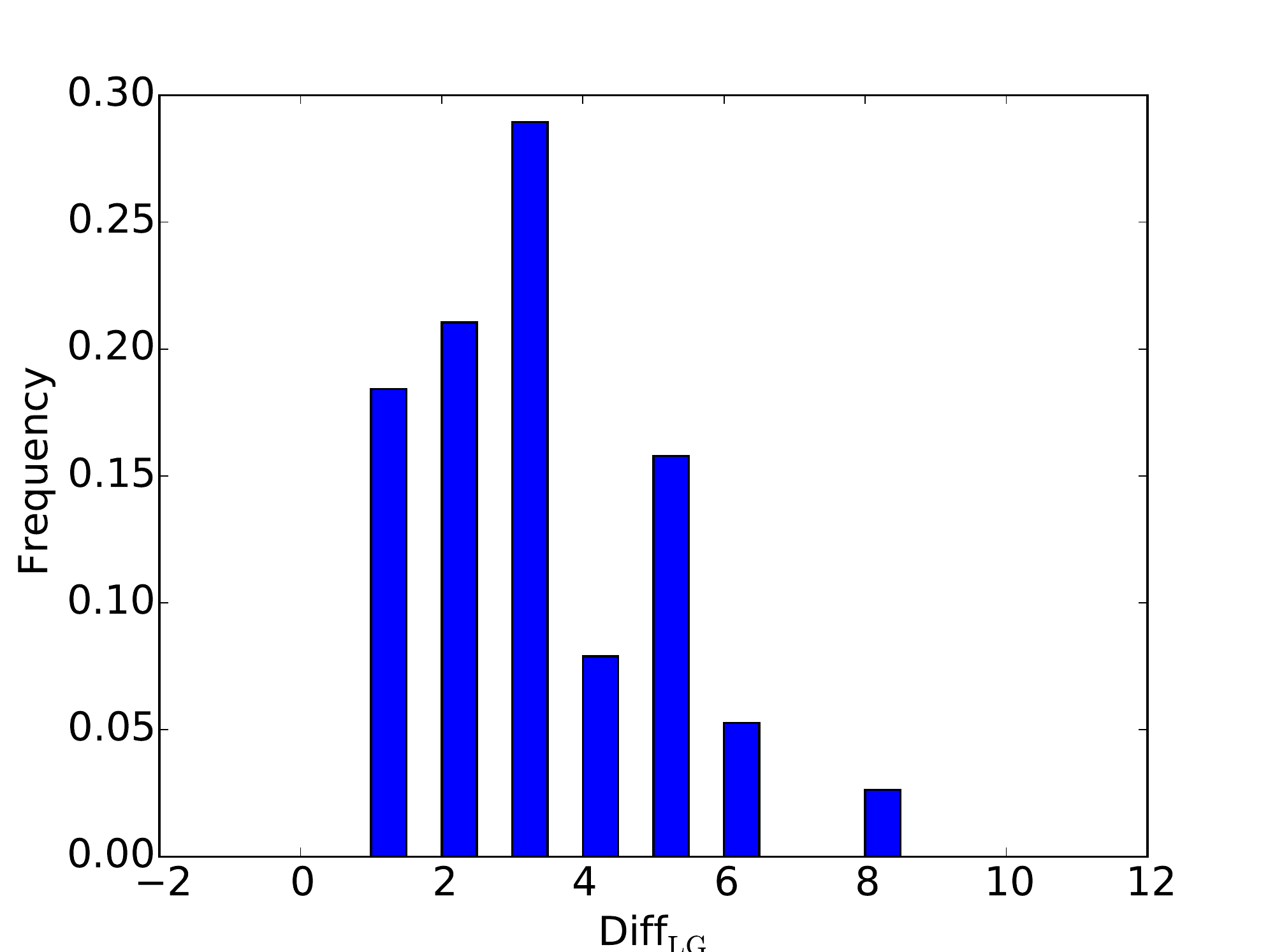}
				\label{subfig:e_distri_cifar10}
			}
			
			\subfloat[CIFAR-100]
			{
				\includegraphics[width=\textwidth]{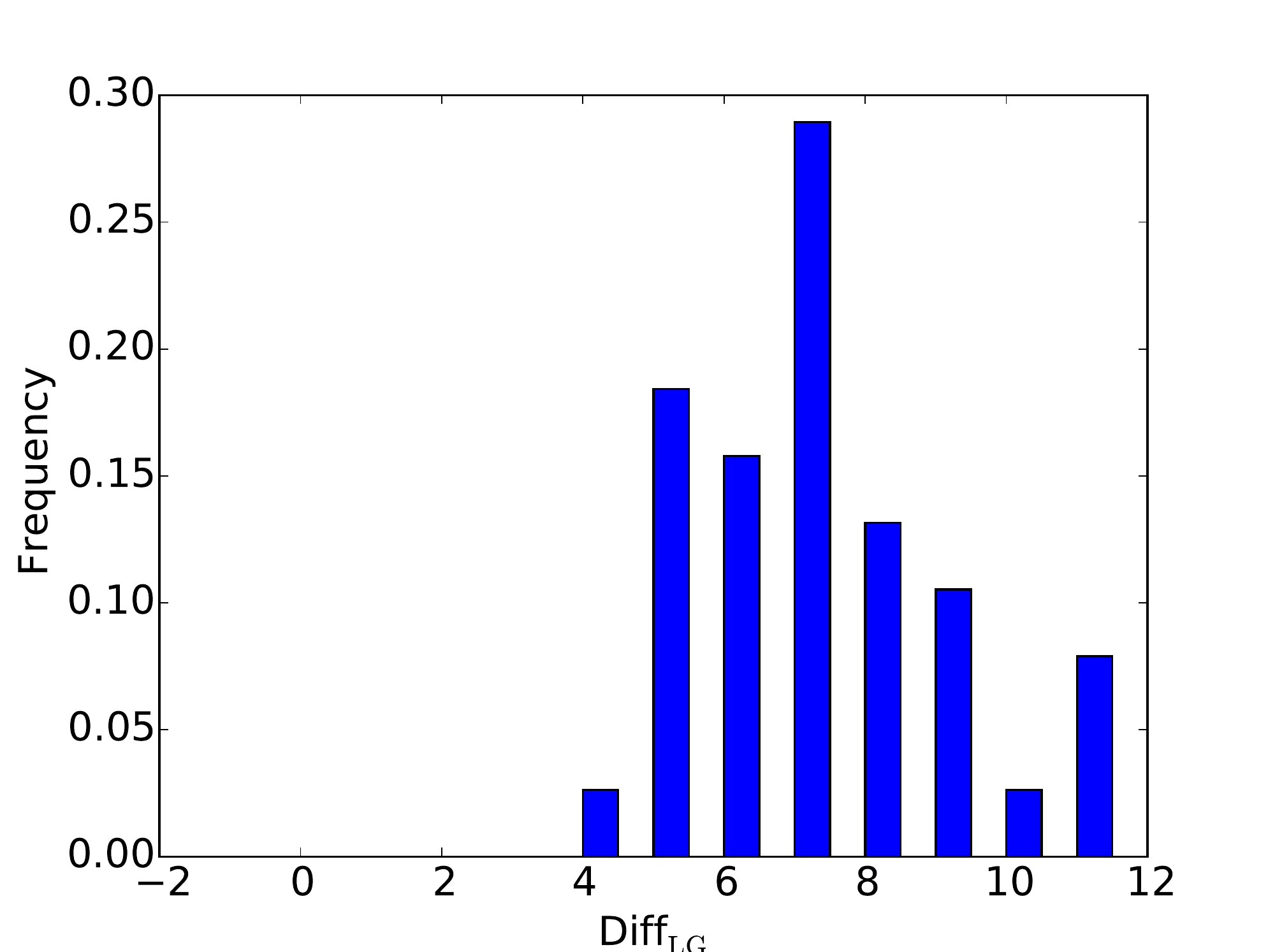}
				\label{subfig:e_distri_cifar100}
			}
			\caption{Ensemble.}
			\label{fig:e_distri}
		\end{minipage}
		\begin{minipage}{0.32\textwidth}
			\centering
			\subfloat[CIFAR-10]
			{
				\includegraphics[width=\textwidth]{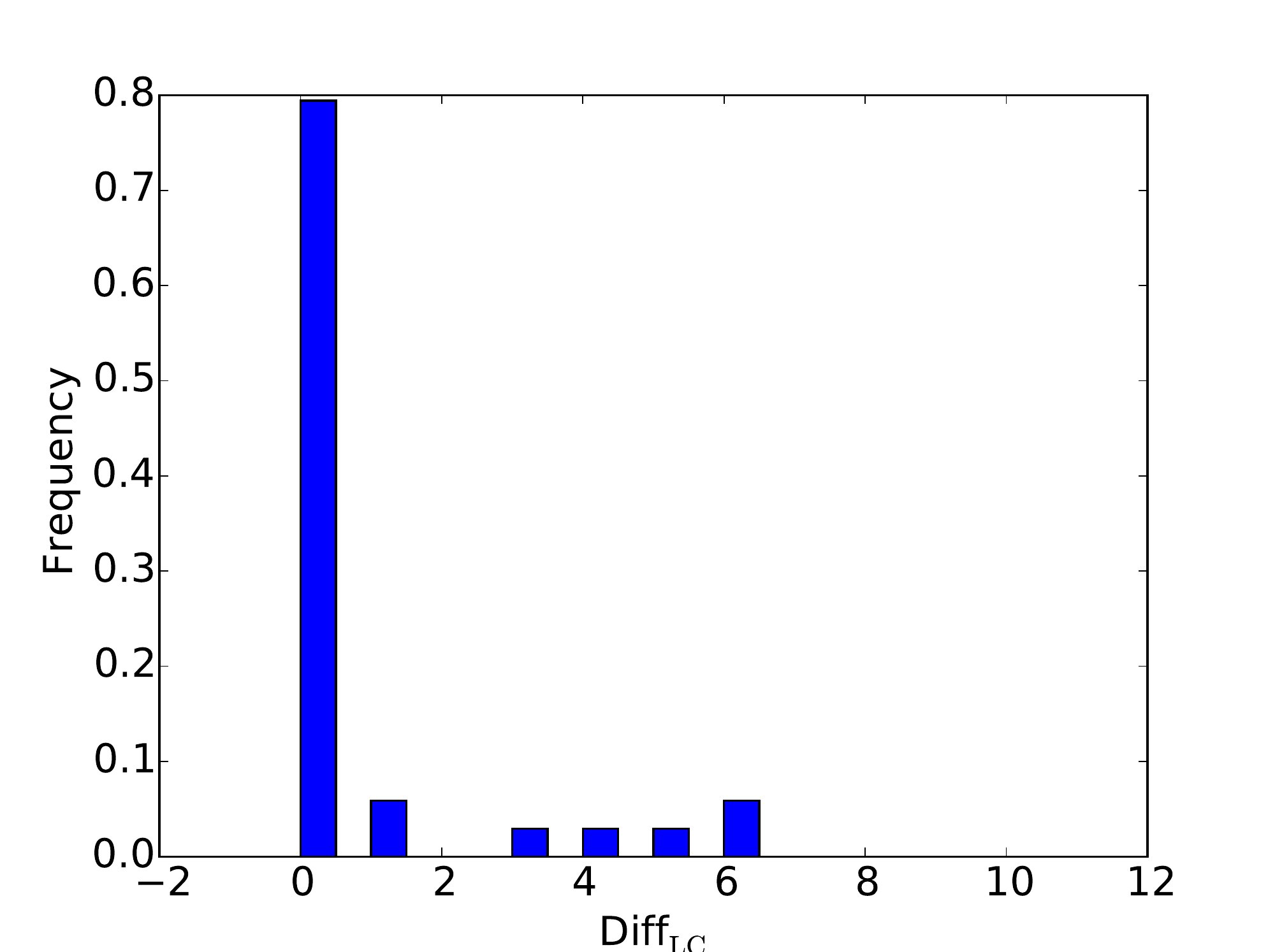}
				\label{subfig:c_distri_cifar10}
			}
			
			\subfloat[CIFAR-100]
			{
				\includegraphics[width=\textwidth]{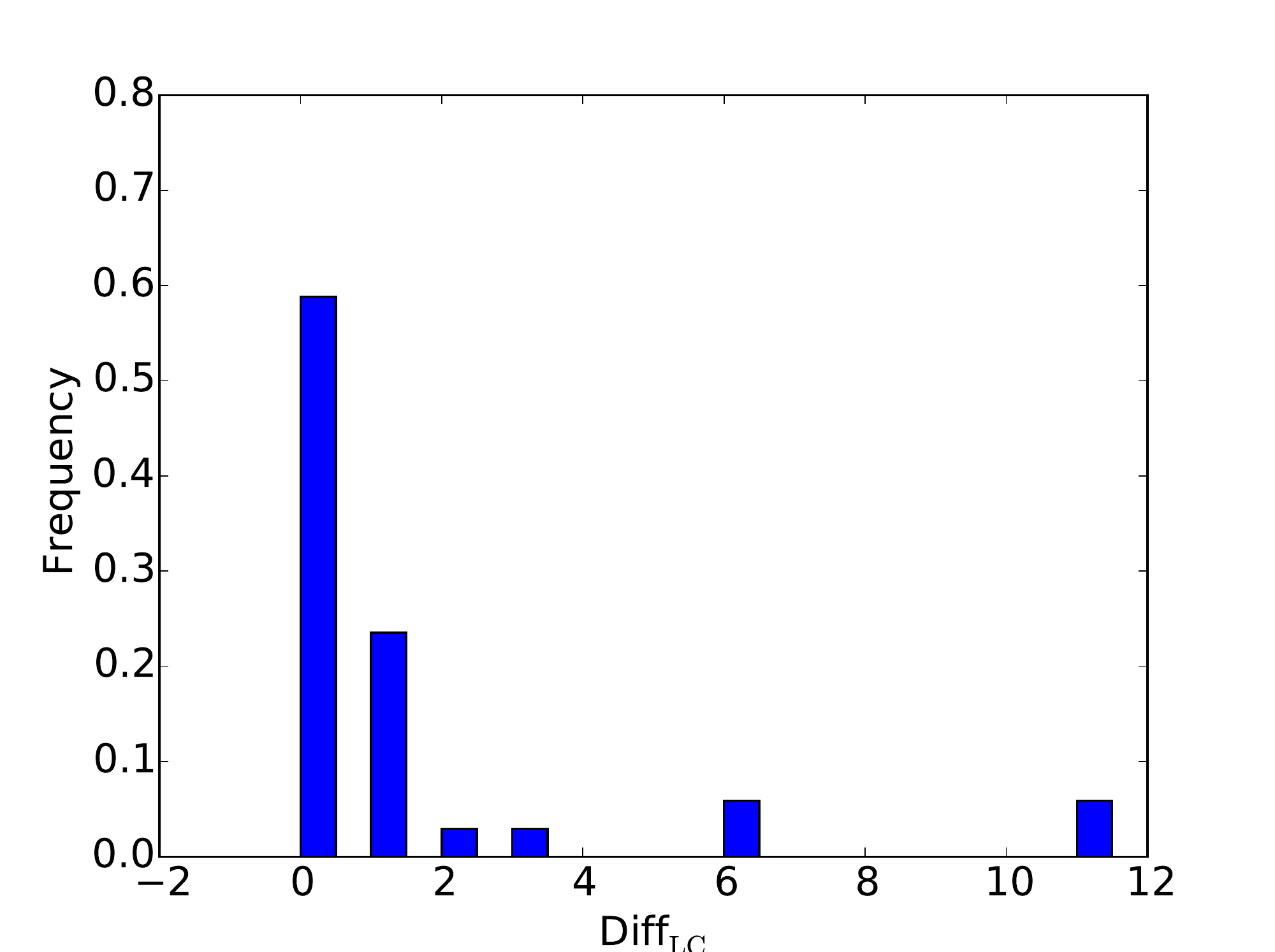}
				\label{subfig:c_distri_cifar100}
			}
			\caption{Compression.}
			\label{fig:c_distri}
		\end{minipage}
	\end{figure}

	We first compare the performance of aggregation methods, i.e. MA and Ensemble. We employ Diff$_\mathrm{LG}$ as the evaluation metric, which measures the improvement of the test error of the global model compared to that of the local models, i.e.,
	\begin{small}
		\begin{equation}\label{eqn:Diff_LG}
		\mathrm{Diff}_\mathrm{LG}= \frac{1}{K}\sum_{k=1}^{K} \mathrm{error}_k-\mathrm{error}_\mathrm{global},
		\end{equation}
	\end{small}where $\mathrm{error}_k$ denotes the test error of the local model on worker $k$, and $\mathrm{error}_\mathrm{global}$ denotes the test error of the corresponding global model produced by MA (or ensemble) in MA-DNN (or EC-DNN). The positive (or negative) Diff$_\mathrm{LG}$ means performance improvement (or drop) of global models over local models. On each dataset, we produce a distribution for Diff$_\mathrm{LG}$ over all the communications and all the parallel settings (including numbers of workers and communication frequencies). We show the distribution for Diff$_\mathrm{LG}$ of MA and ensemble on CIFAR datasets in Fig.~\ref{fig:ma_distri} and \ref{fig:e_distri} respectively, in which red bars (or blue bars) stand for that the performance of the global model is worse (or better) than the average performance of local models.

	For MA, from Fig.~\ref{fig:ma_distri}, we can observe that, on both datasets, over $10\%$ global models achieve worse performance than the average performance of local models, and the average performance of locals model can be worse than the global model by a large margin, e.g., $30\%$. On the other hand, for ensemble, we can observe from Fig.~\ref{fig:e_distri} that the performance of the global model is consistently better than the average performance of the local models on both datasets. Specifically, the performances of over $20\%$ global models are $5$+\% better than the average performance of local models on both datasets.
	
	\subsubsection{Model Compression.}
	In order to avoid model size explosion, the ensemble model is compressed before the next round of local training. However, such compression may result in a risk of performance loss. To examine if the performance improvement of the compressed global models over those local ones in EC-DNN can still outperform such kind of improvement in MA-DNN,  we compare Diff$_\mathrm{LG}$ in MA-DNN (see Eq.(\ref{eqn:Diff_LG})) with Diff$_\mathrm{LC}$ in EC-DNN,
	\begin{small}
		\begin{equation}\label{eqn:Diff_LC}
			\mathrm{Diff}_\mathrm{LC}= \frac{1}{K}\sum_{k=1}^{K} \left(\mathrm{error}_k-\mathrm{error}_{\mathrm{compress},k}\right),
		\end{equation}
	\end{small}where $\mathrm{error}_k$ denotes the test error of the local model on worker $k$, and $\mathrm{error}_{\mathrm{compress},k}$ denotes the test error of the corresponding compressed model after compressing the ensemble model of those local models on worker $k$. The positive (or negative) Diff$_\mathrm{LC}$ means performance improvement (or drop) of the compressed model over local ones. Figure~\ref{fig:c_distri} illustrates the distribution of Diff$_\mathrm{LC}$ over all the communications and various settings of communication frequency and the number of workers on two CIFAR datasets.
	
	From Fig.~\ref{fig:c_distri}, we can observe that the average performance of the compressed models is consistently better than that of the local models on both datasets in EC-DNN, while Figure~\ref{fig:ma_distri} indicates that there are over 10\% global models do not reaching better performance than the local ones in MA-DNN. In addition, the average improvement of compressed models over local ones in EC-DNN is greater than that in MA-DNN. Specifically, the average of such improvements in EC-DNN are 1.03\% and 1.95\% on CIFAR-10 and CIFAR-100, respectively, while the average performance difference in MA-DNN are -3.53\% and 1.72\% on CIFAR-10 and CIFAR-100 respectively.  All these results can indicate that EC-DNN is a superior method than MA-DNN.

	\subsubsection{Accuracy.}
\begin{figure}[t]
		\centering
		\subfloat[$K=4$, CIFAR-10]
		{
			\includegraphics[width=0.5\columnwidth]{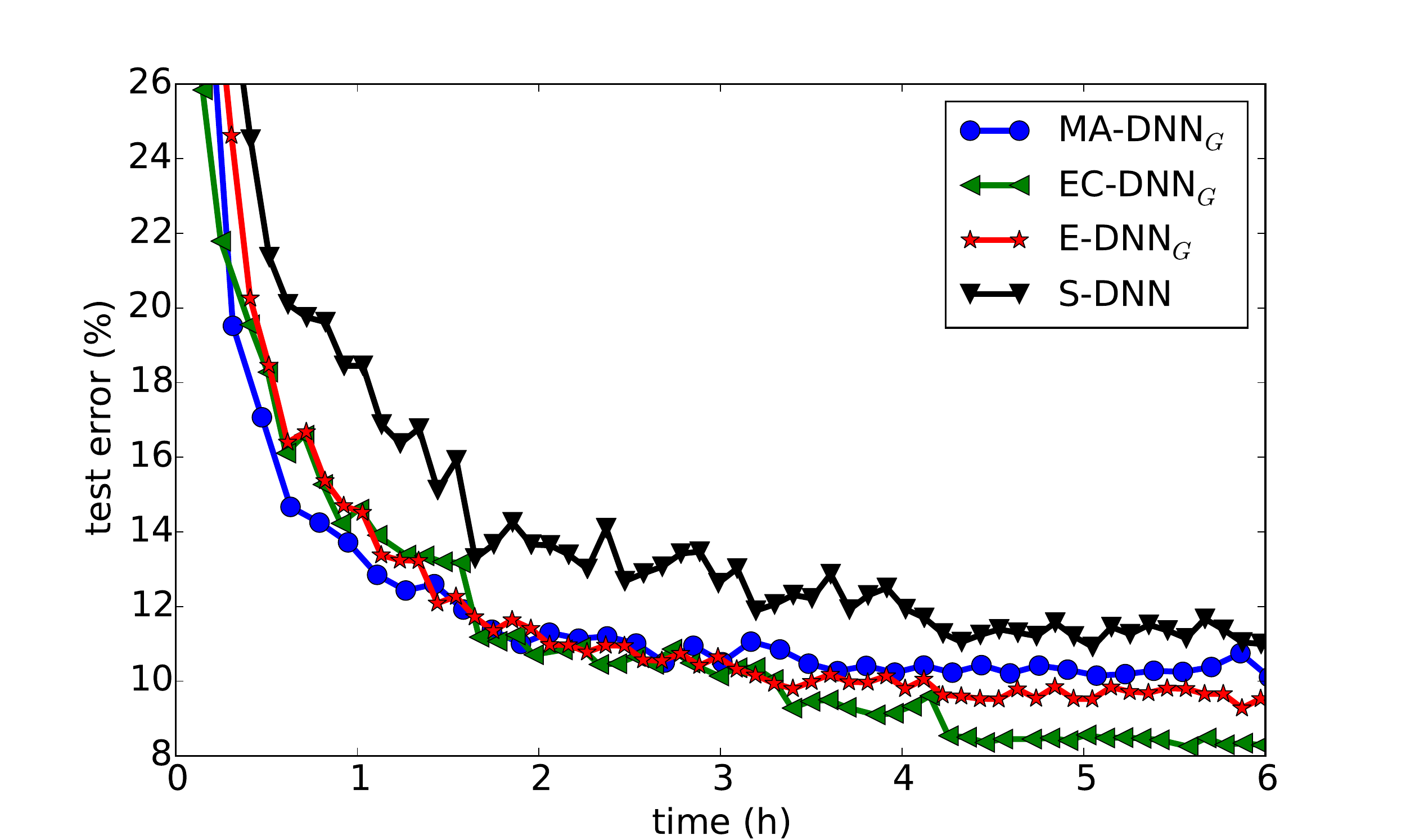}
			\label{subfig:cifar10_ec_4}
		}
		\subfloat[$K=8$, CIFAR-10]
		{
			\includegraphics[width=0.5\columnwidth]{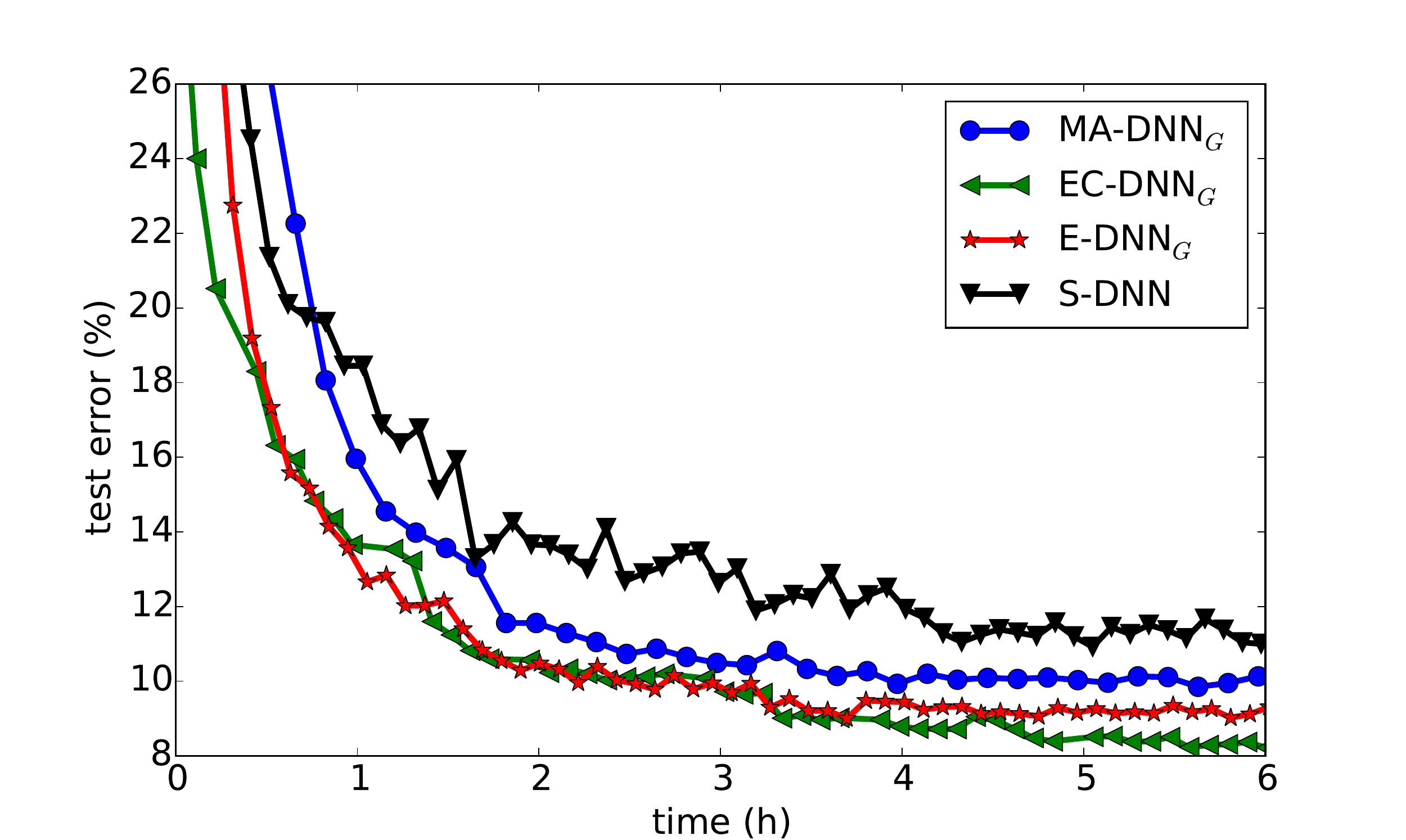}
			\label{subfig:cifar10_ec_8}
		}
				
		\subfloat[$K=4$, CIFAR-100]
		{
			\includegraphics[width=0.5\columnwidth]{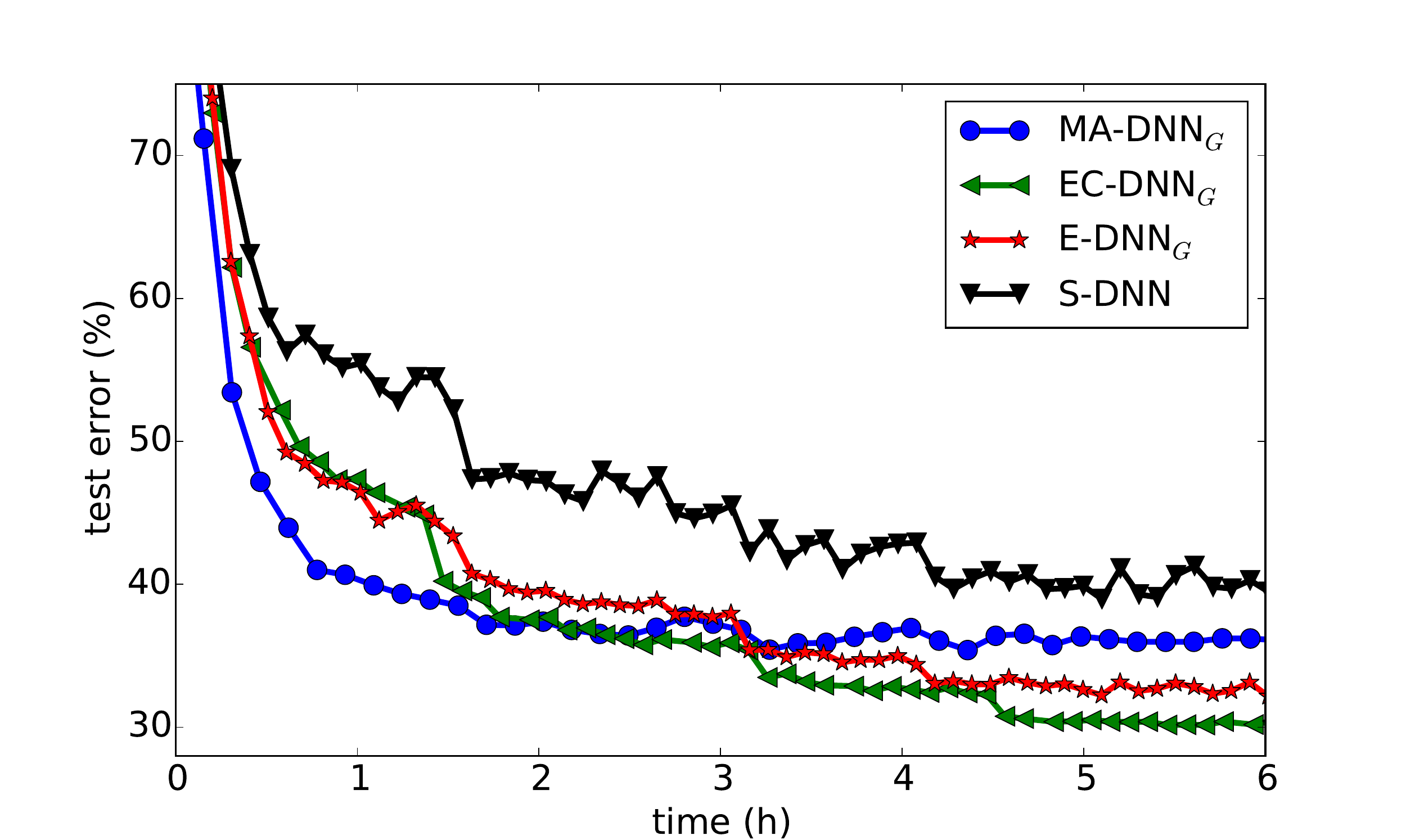}
			\label{subfig:cifar100_ec_4}
		}		
		\subfloat[$K=8$, CIFAR-100]
		{
			\includegraphics[width=0.5\columnwidth]{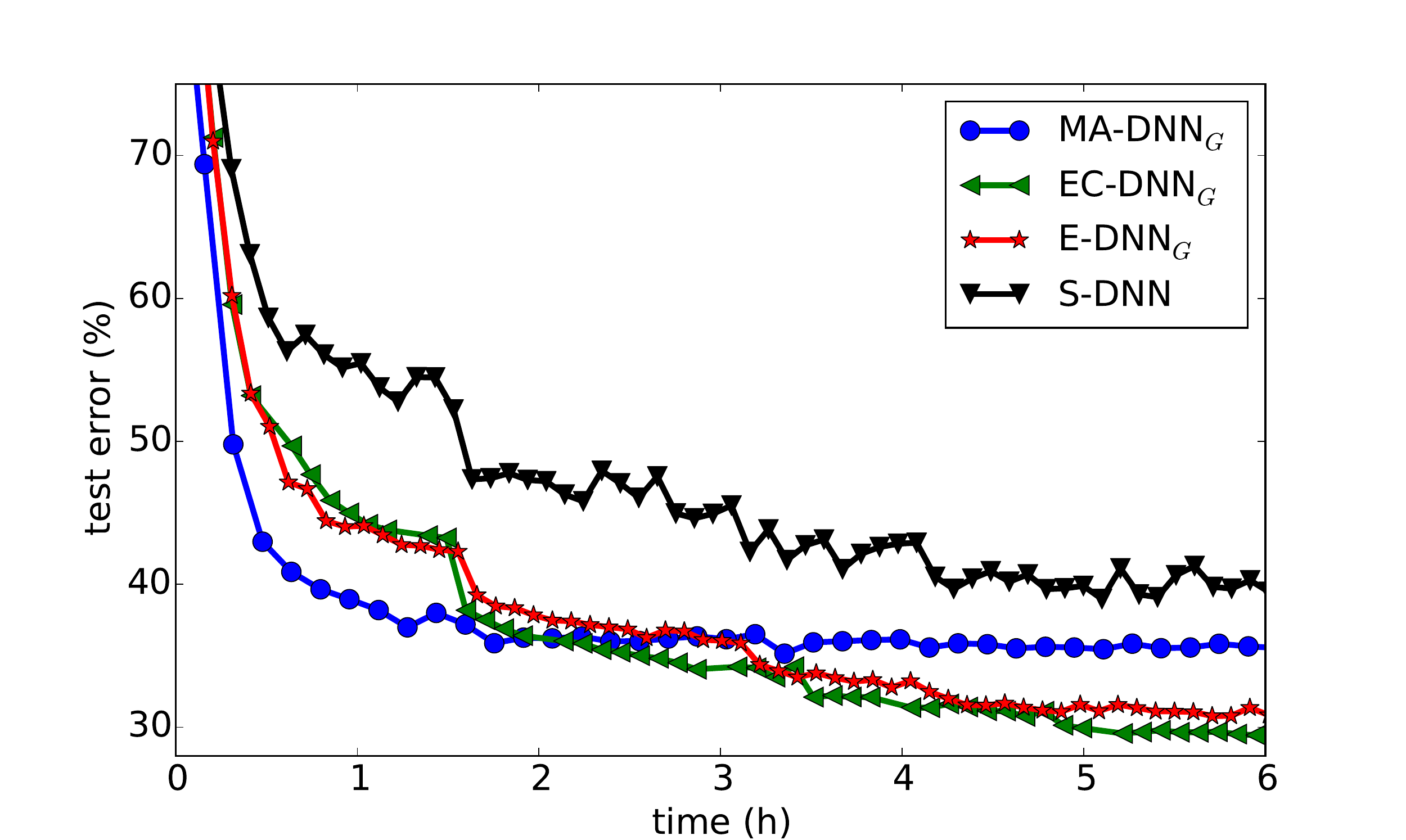}
			\label{subfig:cifar100_ec_8}
		}
		\caption{Test error curves on CIFAR datasets.}
		\label{fig:psgd_ec}
	\end{figure}

	In the following, we examine the accuracy of compared methods. Figure~\ref{fig:psgd_ec} shows the test error of the global model during the training process w.r.t. the overall time, and Table~\ref{tab:test_error} reports the final performance after the training process converges.
	For EC-DNN, the relabeling time has been counted in the overall time when plotting the figure and the table. We report EC-DNN and MA-DNN that achieve best test performance among all the communication frequencies.
	
	From Fig.~\ref{fig:psgd_ec}, we can observe that EC-DNN$_G$ outperforms MA-DNN$_G$ and S-DNN on both datasets for all the number of workers, which demonstrates that EC-DNN is superior to MA-DNN. Specifically, at the early stage of training, EC-DNN$_G$ may not outperform MA-DNN$_G$. We hypothesize the reason as the very limited number of communications among local works at the early stage of EC-DNN training. Along with increasing rounds of communications, EC-DNN will catch up with and then keep outperforming MA-DNN after the certain time slot. Besides, EC-DNN$_G$ outperforms E-DNN$_G$ consistently for different datasets and number of workers, indicating that technologies in EC-DNN are not trivial improvements of E-DNN but is the key factor of the success of EC-DNN.  
	
	In Table~\ref{tab:test_error}, each EC-DNN$_G$ outperforms MA-DNN$_L$ and MA-DNN$_G$. The average improvements of EC-DNN$_G$ over MA-DNN$_L$ and MA-DNN$_G$ are around 1\% and 5\% for CIFAR-10 and CIFAR-100 respectively. Besides, we also report the final performance of EC-DNN$_L$ considering that it can save test time and still outperform both MA-DNN$_L$ and MA-DNN$_G$ when we do not have enough computational and storage resource. Specifically, the best EC-DNN$_L$ achieved test errors of 10.04\% and 9.88\% for $K=4$ and $K=8$ respectively on CIFAR-10, while it achieved test errors of 34.8\% and 35.1\% for $K=4$ and $K=8$ respectively on CIFAR-100. In addition, E-DNN$_L$ never outperforms MA-DNN$_L$ and MA-DNN$_G$.
	
	\subsubsection{Speed.}
	According to our analysis in Sec~\ref{subsec:time}, EC-DNN is more time-efficient than MA-DNN because it communicates less frequently than MA-DNN and thus costs less time on communication. To verify this, we measure the overall time cost by each method to achieve the same accuracy. Table~\ref{tab:test_error} shows the speed of compared methods. In this table, we denote the speed of MA-DNN$_G$ as 1, and normalize the speed of other methods by dividing that of MA-DNN$_G$. If one method never achieves the same performance with MA-DNN$_G$, we denote its speed as 0. Therefore, larger value of speed indicates better speedup. 
	
	From Table~\ref{tab:test_error}, we can observe that EC-DNN can achieve better speedup than MA-DNN on all the datasets. On average, EC-DNN$_G$ and EC-DNN$_L$ runs about 2.24 and 1.33 times faster than MA-DNN$_G$, respectively. Furthermore, EC-DNN consistently results in better speedup than E-DNN on all the datasets. On average, E-DNN$_G$ only runs about 1.85 times faster than MA-DNN$_G$ while EC-DNN$_G$ can reach about 2.24 times faster speed. From this table, we can also find that E-DNN$_L$ never achieves the same performance with MA-DNN$_G$ while EC-DNN$_L$ can contrarily run much faster than MA-DNN$_G$.
	
	Furthermore, Table~\ref{tab:test_error} demonstrates the communication frequency $\tau$ that makes compared methods achieve the corresponding speed. We can observe that EC-DNN tend to communicate less frequently than MA-DNN. Specifically, MA-DNN usually achieves the best performance with a small $\tau$ (i.e., 16), while EC-DNN cannot reach its best performance before $\tau$ is not as large as 2000. 
		
	\begin{table}[t]
		\centering
		\caption{Test error (\%), speed and $\tau$ on CIFAR datasets.}\label{tab:test_error}
		\begin{small}
			\begin{tabular}{c|ccc|ccc||ccc|ccc}
				\hline
				\hline
				& \multicolumn{6}{c||}{CIFAR-10} & \multicolumn{6}{c}{CIFAR-100} \\
				\hline
				& \multicolumn{3}{c|}{K=4} &\multicolumn{3}{c||}{K=8} &\multicolumn{3}{c|}{K=4} & \multicolumn{3}{c}{K=8}\\
				\hline
				& Error &  Speed &$\tau$ & Error &  Speed & $\tau$ & Error &  Speed & $\tau$ & Error &  Speed &$\tau$ \\
				\hline
				MA-DNN$_G$ & 10.3 &  1 & 16 & 9.99 &  1 & 2k & 36.18 &  1 & 16 & 35.55  & 1 & 16\\
				E-DNN$_G$ & 9.44 &  1.58 & - & 9.05 &  1.92 & - & 32.49 &  1.95 & - & 30.9 &  1.97 & - \\
				EC-DNN$_G$ & \textbf{8.43} &  \textbf{1.92} & 4k & \textbf{8.19} &  \textbf{2.05} & 4k & \textbf{30.26} & \textbf{2.52} & 4k & \textbf{29.31} &  \textbf{2.48} & 2k \\
				\hline	
				MA-DNN$_L$ & 10.55&  0 & 16 & 10.54 &  0 & 2k & 36.39 &  0 & 16 & 35.56 &  0 & 16 \\
				E-DNN$_L$ & 11.04&  0 & - & 10.95 & 0 & - & 39.57 & 0 & - & 39.55 & 0 & - \\
				EC-DNN$_L$ & \textbf{10.04}& \textbf{1.36} & 4k & \textbf{9.88}  & \textbf{1.26} & 4k & \textbf{34.8} & \textbf{1.42} & 4k & \textbf{35.1} & \textbf{1.27} & 2k \\
				\hline
				S-DNN & \multicolumn{6}{c||}{10.41} & \multicolumn{6}{c}{35.68}\\
				\hline
				\hline
			\end{tabular}
		\end{small}
	\end{table}

	\subsubsection{Large-Scale Experiments.}
\begin{figure}[t]
		\centering
		\subfloat[$K=4$]
		{
			\includegraphics[width=0.5\linewidth]{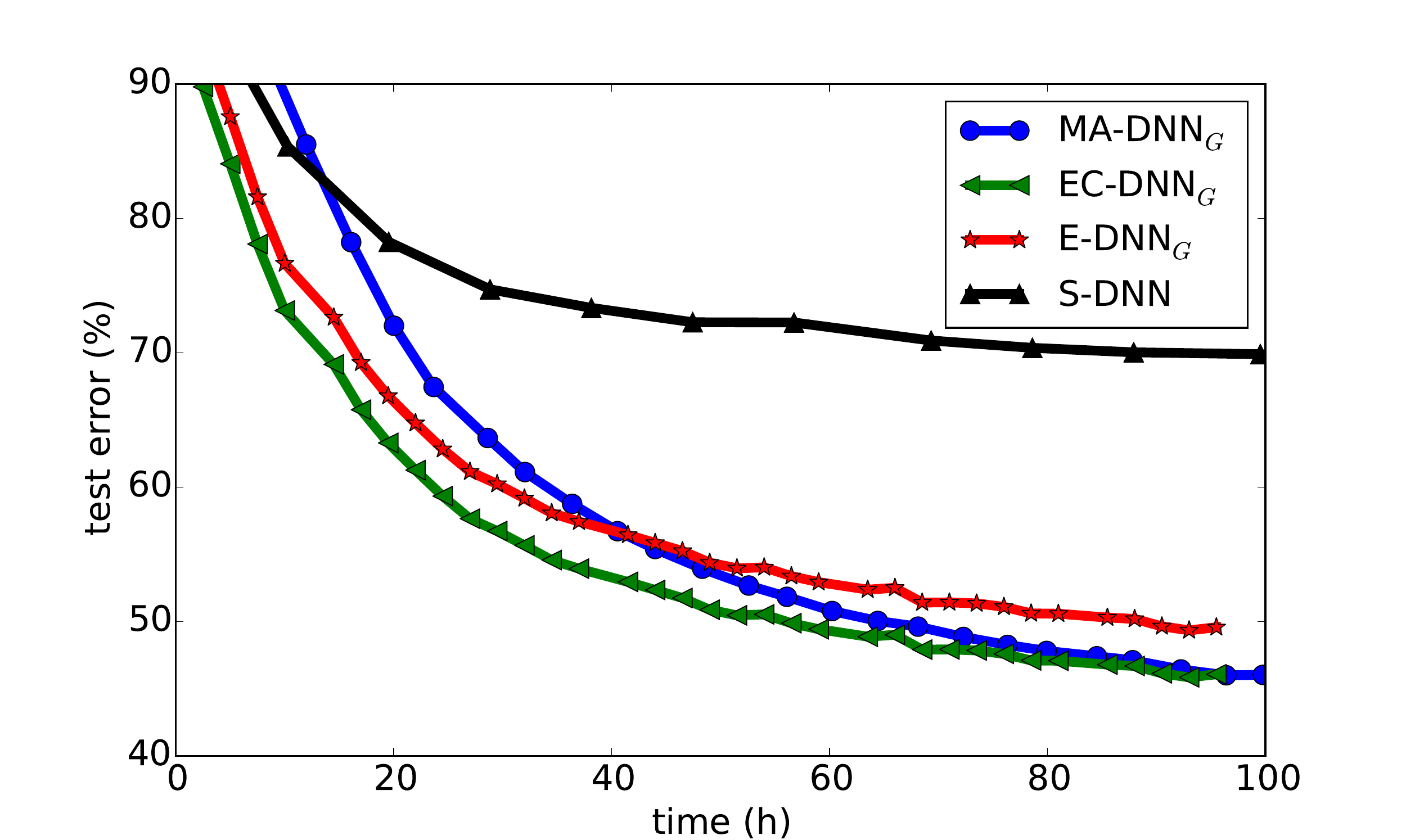}
			\label{subfig:imagenet_ec_4}
		}
		\subfloat[$K=8$]
		{
			\includegraphics[width=0.5\columnwidth]{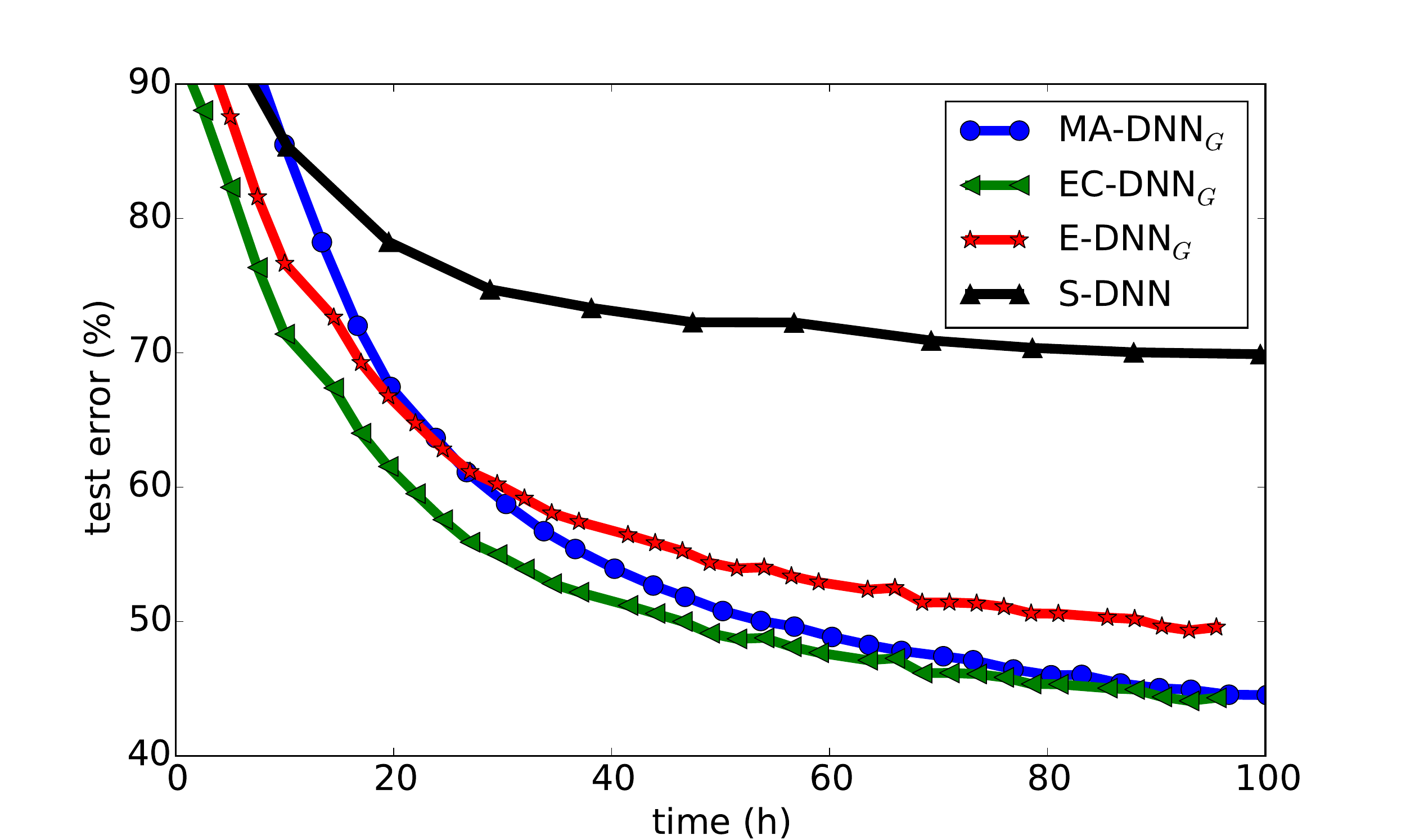}
			\label{subfig:imagenet_ec_8}
		}	
		\caption{Test error curves on ImageNet.}
		\label{fig:imagenet_ec_ima}
	\end{figure}

	In the following, we will conduct experiments to compare the performance of MA-DNN with that of EC-DNN with the setting of much bigger model and more data, i.e., GoogleNet on ImageNet. Figure~\ref{fig:imagenet_ec_ima} shows the test error of the global model w.r.t the overall time. The communication frequencies $\tau$ that makes MA-DNN and EC-DNN achieve best performance are 1 and 1000 respectively. We can observe that EC-DNN consistently achieves better test performance than S-DNN, MA-DNN and E-DNN throughout the training. Besiedes, we can observe that EC-DNN outperforms MA-DNN even at the early stage of the training, while EC-DNN cannot achieve this on CIFAR datasets because it communicates less frequently than MA-DNN. The reason is that frequent communication will make the training much slower for very big model, i.e., use less mini-batches of data within the same time. When the improvements introduced by MA cannot compensate the decrease of the number of used data, MA-DNN no longer outperforms EC-DNN at the early stage of the training. In this case, the advantage of EC-DNN becomes even more outstanding. 
	
	\section{Conclusion and Future Work}\label{sec:discussion}
	In this paper, we propose EC-DNN, a new Ensemble-Compression based parallel training framework for DNN. As compared to the traditional approach, MA-DNN, that averages the parameters of different local models, our proposed method uses the ensemble method to aggregate local models. In this way, we can guarantee that the error of the global model in EC-DNN is upper bounded by the average error of the local models and can consistently achieve better performance than MA-DNN. In the future, we plan to consider other compression methods for EC-DNN. Besides, we will investigate the theoretical properties of the ensemble method, compression method, and the whole EC-DNN framework.
	\section{Acknowledgments}
	This work is partially supported by NSF of China (grant numbers: 61373018, 61602266, 11550110491) and  NSF of Tianjin (grant number: 4117JCYBJC15300). 
	\bibliographystyle{splncs03}
	\bibliography{ref}
\end{document}